 \documentclass[preprint ]{aastex} % preprint   (1-column, single-spaced)
\shorttitle{Mauche \& Raymond}
\shortauthors{EUVE Observations of OY Car}

 \slugcomment{Accepted for publication in
             {\it     Astrophysical Journal\/} April 27, 2000}
\newcommand\Mwd   {M_{\rm wd}}
\newcommand\Rwd   {R_{\rm wd}}
\newcommand\Msun  {{\rm M_{\odot}}}
\newcommand\Lsun  {{\rm L_{\odot}}}
\newcommand\Mdot  {\dot{M}}
\newcommand\Mdota {\dot{M}_{\rm a}}
\newcommand\Mdotw {\dot{M}_{\rm w}}
\newcommand\Pdot  {\dot{P}}
\newcommand\lax{{\lower0.75ex\hbox{ $<$ }\atop\raise0.5ex\hbox{ $\sim$ }}}
\newcommand\gax{{\lower0.75ex\hbox{ $>$ }\atop\raise0.5ex\hbox{ $\sim$ }}}

\begin{document}

\title{EUVE Observations of OY Carinae in Superoutburst}

\author{Christopher W.\ Mauche}
\affil{Lawrence Livermore National Laboratory, \\
       L-43, 7000 East Avenue, Livermore, CA 94550; \\
       mauche@cygnus.llnl.gov}
\and
\author{John C.\ Raymond}
\affil{Harvard-Smithsonian Center for Astrophysics, \\
       60 Garden Street, Cambridge, MA 02138; \\
       jraymond@cfa.harvard.edu}

\clearpage % force page break; deleted when you change from pp to ms

% Abstract
%---------------------------------------------------------

\begin{abstract}

The {\it Extreme Ultraviolet Explorer\/} ({\it EUVE\/}) satellite was
used for three days beginning on 1997 Mar 26.96 UT to obtain photometric
and spectroscopic observations of the eclipsing SU~UMa-type dwarf nova
OY~Carinae in superoutburst. Because of the longer time on source (143
ks), the larger number of eclipses observed (17), and the higher count
rate in the detector (0.5--2.2 $\rm counts~s^{-1}$), we are able to
significantly strengthen previous reports that there is little or no
eclipse by the secondary of the EUV emission region of OY~Car in
superoutburst. The mean {\it EUVE\/} spectrum extends from 70 to
190~\AA \ and contains broad ($\rm FWHM\approx 1$~\AA ) emission lines
of \ion{N}{5}, \ion{O}{5}--\ion{O}{6}, \ion{Ne}{5}--\ion{Ne}{7},
\ion{Mg}{4}--\ion{Mg}{6}, \ion{Fe}{6}--\ion{Fe}{8}, and possibly
\ion{Fe}{23}. Good fits of the observed spectrum are obtained with a
model (similar to that of Seyfert 2 galaxies) wherein radiation from
the boundary layer and accretion disk is scattered into the line of
sight by the system's photoionized accretion disk wind. It is possible
to trade off continuum luminosity for wind optical depth, but reasonable
models have a boundary layer temperature $T_{\rm bl}\approx 90$--130~kK
and a boundary layer and accretion disk luminosity $L_{\rm bl}=
L_{\rm disk}\lax 4\times 10^{34}~\rm erg~s^{-1}\approx 10\> \Lsun $,
corresponding to a mass-accretion rate $\Mdota\lax 10^{-8}~\rm
\Msun~yr^{-1}$; an absorbing column density $N_{\rm H}\approx
1.6$--$3.5\times 10^{19}~\rm cm^{-2}$; and a wind mass-loss rate
$\Mdotw\lax 10^{-10}~{\rm \Msun~yr^{-1}} \approx 0.01\, \Mdota $.
Because radiation pressure alone falls an order of magnitude short of
driving such a wind, magnetic forces must also play a role in driving
the wind of OY~Car in superoutburst.

\end{abstract}


\clearpage % force page break; delete when you change from pp to ms

% Body of the paper
%---------------------------------------------------------

\section{Introduction}

Cataclysmic variables (CVs) are a diverse class of semidetached binaries
composed typically of a low-mass main-sequence secondary, an accreting
white dwarf, and a luminous accretion disk. There are three generalized
classes of CVs: novae, novalike variables, and dwarf novae. In dwarf
novae, a mass-transfer instability causes the disk to brighten by 3--5
magnitudes in the optical at quasi-periodic intervals of tens to hundreds
of days. Based on their light curve morphologies, most dwarf novae are
classified as either U~Gem systems with ``normal'' outbursts only, Z~Cam
systems with ``normal'' outbursts and prolonged states of intermediate
brightness (``standstills''), or SU~UMa systems with ``normal'' and
``super'' (brighter, longer-duration) outbursts. Nonmagnetic novalike
variables behave like dwarf novae ``stuck'' in outburst. \citet{war95}
provides an excellent background on all of the various CV types and
subtypes, while \citet{can93} and \citet{osa96} provide recent reviews of
the theory of dwarf nova outbursts.

While the white dwarf and accretion disk are the dominant sources of
optical through FUV light in dwarf novae, the boundary layer between
the disk and the surface of the white dwarf is the dominant source of
higher-energy emission. Modulo a correction factor for white dwarf
rotation, the boundary layer luminosity $L_{\rm bl}= G\Mwd\Mdota/
2\Rwd\sim 4\times 10^{34}\, (\Mdota/10^{-8}~\rm
\Msun~yr^{-1})~erg~s^{-1}$, where $\Mdota $ is the accretion rate and
$\Mwd $ and $\Rwd $ are respectively the mass and radius of the white
dwarf. When $\Mdota $ is low (e.g., dwarf novae in quiescence), the
boundary layer is optically thin and quite hot (of order the virial
temperature $T_{\rm vir} = G\Mwd m_{\rm H}/3k\Rwd\sim {\rm tens}$ of keV;
\citealt{pri79, tyl81, nar93}); when $\Mdota $ is high (e.g., novalike
variables and dwarf novae in outburst), the boundary layer is optically
thick and quite cool (of order the blackbody temperature $T_{\rm bb}
=[G\Mwd\Mdota/8\pi\sigma\Rwd ^3]^{1/4} \sim 100$ kK; \citealt{pri77,
tyl77, pop95}).

Observationally, the location and extent of the source of the X-rays in
nonmagnetic CVs is constrained by (1) the X-ray light curves of eclipsing
systems, (2) the variation of apparent emission measure with binary
inclination \citep{tes96}, and (3) the relative strength of the line and
continuum spectrum reflected off the white dwarf and accretion disk
\citep[e.g.,][]{don97}. All these constraints imply that all or most of
the hard X-rays in low-$\Mdot $ CVs are emitted very close to the white
dwarf, but the eclipse observations provide the most direct and
unambiguous evidence. In HT~Cas ($i\approx 81^\circ $; \citealt{woo95a,
muk97}), Z~Cha ($i\approx 82^\circ $; \citealt{tes97}), and OY~Car
($i\approx 83^\circ $; \citealt{pra99a}) in quiescence, the X-rays are
fully eclipsed for an orbital phase interval comparable to that of the
optical eclipse of the white dwarf.

In contrast to the compact source of hard X-rays in low-$\Mdot $ CVs,
there is strong evidence for an extended source of soft X-rays in
high-$\Mdot $ CVs. {\it EXOSAT\/} and {\it ROSAT\/} light curves of
OY~Car in superoutburst \citep{nay88, pra99b} and {\it ROSAT\/} light
curves of the novalike variable UX~UMa ($i\approx 71^\circ $;
\citealt{woo95b}) rule out an eclipse by the secondary of a compact
source of X-rays centered on the white dwarf. In both cases, it is argued
that the boundary layer is obscured from view by the accretion disk at
all orbital phases, and that the observed X-rays come from a corona or
wind above the disk.

The situation is far more complex in U~Gem ($i\approx 70^\circ $). In
quiescence, the hard X-rays are partially eclipsed at orbital phase $\phi
\sim 0.7$ \citep{szk96}, and during outburst the EUV and soft X-rays are
partially eclipsed at orbital phases $\phi\sim 0.7$ and 0.1 \citep{mas88,
lon96}. These dips in the X-ray and EUV light curves are interpreted as
partial eclipses of the boundary layer by vertical structure at the edge
of the accretion disk ($r\approx 4\times 10^{10}$~cm) or at the
circularization radius of the accretion stream ($r\approx 5\times
10^9$~cm); in either case, the heights above the orbital plane are
distressingly large: $h\approx 0.5\, r$, whereas the disk thickness
$H\approx c_{\rm s}/\Omega_{\rm K}\ll r$. The {\it EUVE\/} spectra of
U~Gem in outburst supply additional diagnostic information. The
phase-averaged spectrum contains emission lines of Ne VI--VIII, Mg
VI--VII, and Fe VII--X superposed on a $T= 110$--140~kK blackbody
continuum \citep{lon96}. The phase-resolved spectra demonstrate that the
eclipses affect the continuum more strongly than the lines (most
strikingly, the Ne~VIII $\lambda 88.1$ line is nearly unaffected by the
eclipses; see \citealt{mau97}), implying that the lines are produced in a
region of larger extent than that of the continuum, which presumably is
formed in the boundary layer.

To gain a better understanding of the extended source of soft X-rays
in high-$\Mdot $ CVs, we obtained {\it EUVE\/} observations of the
eclipsing SU~UMa-type dwarf nova OY~Car in superoutburst. The resulting
data are superior to those from the previous {\it EXOSAT\/} and {\it
ROSAT\/} observations because of the longer time on source, the larger
number of eclipses observed, the higher count rate in the detector, and
the ability to obtain dispersed EUV spectra. We discuss the {\it EUVE\/}
observations in \S 2, the DS light curve in \S 3, and the SW spectrum
in \S 4. We find that there is little or no eclipse of the EUV emission
region in OY~Car in superoutburst (\S 3), and that the EUV spectrum
contains emission lines of intermediate ionization stages of N, O, Ne,
Mg, and Fe (\S 4). The spectrum is not fit well in detail by a model of a
collisionally ionized plasma (\S 5), and although it is fit well by a
model wherein radiation from the boundary layer and accretion disk is
scattered into the line of sight by OY~Car's accretion disk wind (\S 6),
radiation pressure alone falls an order of magnitude short of driving the
wind (\S 7). We close in \S 8 with a discussion and summary of our
results.

\section{{\it EUVE\/} Observations}

The events leading up to the {\it EUVE\/} observations of OY~Car are as
follows. On 1997 Mar 25.46 UT M.\ Mattiazzo posted an alert to VSNET
(\#802) and the AAVSO that OY~Car had gone into outburst. On Mar 25.76
this alert was transmitted to us by J.~Mattei, Director of the AAVSO.
Less than an hour after being contacted by us (and despite the fact that
the {\it EUVE\/} Peer Review was ongoing), {\it EUVE\/} Acting Deputy
Director J.\ Vallerga granted our request for target-of-opportunity
observations of OY~Car. On Mar 26.13 M.\ Keane, responding to a request
posted by us to VSNET (\#804), provided independent confirmation of the
outburst in the form of an image of the OY~Car field obtained by N.\
Suntzeff with the 0.9-m telescope at CTIO. In the hours leading up to the
start of {\it EUVE\/} observations on Mar 26.96, photometric estimates
were provided by M.\ Mattiazzo, P.\ Nelson, R.\ Stubbings, T.\ Cragg,
and J.\ Herr.

The optical light curve of the 1997 March/April outburst of OY~Car,
assembled from observations submitted to and obtained from the AAVSO
(J.~A.\ Mattei 1997, personal communication) and VSS/RASNZ (F.~M.\
Bateson 1997, personal communication) is shown in Figure~1. Based on its
peak brightness ($V\approx 11$) and duration ($\Delta t\approx 20$ days),
this event was clearly a superoutburst. According to the records of the
AAVSO, the previous and subsequent two outbursts of OY~Car took place
in 1995 October/November (a superoutburst), 1997 July/August (a normal
outburst), and 1998 February (a superoutburst). For Figure~1, we have
suppressed observations obtained near, and hence affected by, the eclipse
by the secondary of the white dwarf and accretion disk. Specifically, we
assume the binary ephemeris of \citet{pra99a} and retain only those
observations for which the binary phase was in the range $\phi=0.1$--0.9.
The amplitude of the scatter remaining in the light curve is reasonably
consistent with that expected for AAVSO and RASNZ data for such a
relatively dim source.

{\it EUVE\/} observations of OY~Car began on 1997 Mar 26.96 UT, ran for
three days ($\rm JD-2450000=534.457$--537.636), and resulted in a total
of 143 ks of exposure. For a description of the {\it EUVE\/} satellite
and instrumentation, refer to \citet{boy91}, \citet{abb96}, and
\citet{sir97}. It is sufficient to note here that the bandpasses of the
deep survey (DS) photometer and short wavelength (SW) spectrometer are
defined by a Lexan/Boron filter and extend from $\approx 70$ to $\approx
180$~\AA . The medium wavelength (MW) spectrometer bandpass is defined by
an Aluminum/Carbon filter and extends from $\approx 170$ to $\approx
370$~\AA , but interstellar absorption extinguishes the flux longward
of about 200~\AA . These observations were ``spiral dithered'' to avoid
the dead spot near the center of the field of view of the DS instrument,
and to increase the signal-to-noise ratio of the EUV spectrum by averaging
over quantum efficiency variations of the SW detector.

\vspace*{-0.25 in}
\section{DS Light Curve}

The background-subtracted DS count rate light curve from our observations
is shown in Figure~2 and in Figure~1 superposed on the optical light
curve. The data are binned into detector ``good time intervals'' of
typically $\approx 1950$~s, so the error bars are typically smaller than
the points. Given that the optical light curve was constant or declining
only gradually during the {\it EUVE\/} observation, the relatively rapid
decline and subsequent rise of the DS light curve is something of a
surprise. The observation is unfortunately too short to know for sure,
but the similarity to the {\it Voyager\/} FUV light curve of VW~Hyi in
superoutburst \citep{pol87, pri87} suggests that we are seeing in the EUV
the decay of a narrow outburst and the subsequent rise of a long outburst
which ``add'' to produce the superoutburst. This behavior is at least in
qualitative agreement with the thermal-tidal instability model of
superoutbursts \citep{osa96}: a normal outburst is triggered by the disk
thermal instability \citep{can93}, which is turn triggers the tidal
instability when the disk expands out to the 3:1 eccentric inner Lindblad
resonance \citep{lub91}; the resulting enhanced tidal torques enhance the
mass-accretion rate through the disk, rejuvenating the initial outburst.
Credence for this interpretation is supplied by the SPH simulations of
\citet[][Fig.~11]{mur98}, but better (more extensive, multiwavelength)
data are required to convincingly test this interpretation.

One of the purposes of the {\it EUVE\/} observation of OY~Car was a
study of the eclipses (or lack thereof) of the EUV emission region by
the secondary, but this is made difficult by the strong commensurability
between {\it EUVE\/}'s orbital period (94.6 min) and OY~Car's binary
period (90.9 min). The result is that the binary phase advances by only
4\% each satellite orbit, and eclipse intervals ``come around'' only
every 1.6 days. The net result is that during three days of observations
we recorded only 17 eclipses of the white dwarf. The orbits during which
these eclipses occurred are indicated by the circles in Figure~2. During
two of these orbits we are both fortunate and lucky to have a detailed
record of the optical behavior of the source measured by W.~S.~G.\ Walker
(1997, personal communication) with a 10-in telescope and a ST6 CCD
camera. The EUV and optical light curves during the interval of optical
measurements are shown in Figure~3, where the phase is relative to the
binary ephemeris of \citet{pra99a} and the grey stripes denote the
intervals when the white dwarf is eclipsed by the secondary ($-0.02\le
\phi\le +0.02$). Whereas the optical light curve clearly shows the
prominent eclipses of the accretion disk by the secondary, there is no
hint of eclipses in the EUV light curve.

To help quantify this statement, we determined the DS count rate light
curves for each of the intervals containing an eclipse of the white
dwarf. These are shown in the upper panel of Figure~4. To increase the
photon statistics and decrease the effect of the flickering apparent in
the DS light curves, we calculated the mean DS count rate light curves
for the two groups of eight intervals of white dwarf eclipses. Exactly
how this is done is somewhat subjective, but proper account has to made
of the different count rates during each interval $i$ and the different
exposure times within each phase bin $j$. The former was accounted for
by fitting the background-subtracted DS count rate data with a linear
function $f_{ij}=a_i+b_i(\phi_j-\phi_{0,i})$, where $\phi_{0,i}$ is the
midpoint of interval $i$. If $S_{ij}$ and $B_{ij}$ are the counts from
the source and background regions, respectively, and $\Delta t_{ij}$ are
the deadtime and primbsch-corrected exposure times, the mean relative
DS count rate light curves are $(\sum_iS_{ij}-0.2\,\sum_iB_{ij})/
(\sum _i\Delta t_{ij}\, \sum_if_{ij})$, where the factor of 0.2 accounts
for the relative size of the source and background regions. The resulting
two mean relative DS count rate light curves are shown in the lower panels
of Figure~4, where we have retained only those phase bins sampled at
least five times during the eight intervals of white dwarf eclipses. What
one concludes about the depth of the eclipse of the EUV emission region
depends sensitively on the assumed width of the eclipse and the assumed
shape of the uneclipsed light curve. During the first interval the DS
light curve is well behaved and a secure upper limit is 5\% at the 90\%
confidence level. During the second interval the DS light curve is less
well behaved, but a reasonable upper limit is 15\% at the 90\% confidence
level. These results significantly strengthen the conclusion based on
previous {\it EXOSAT\/} \citep{nay88} and {\it ROSAT\/} \citep{pra99b}
observations that there is little or no eclipse by the secondary of the
EUV emission region of OY~Car in superoutburst.

\section{SW Light Curve and Spectrum}

The SW spectrometer provides an independent measurement of the EUV light
curve of OY~Car, but it is less useful than the DS photometer to study
the eclipses because it has approximately one-tenth the effective area
and (because the flux is dispersed over many more detector pixels) a
higher effective background. The upper panel of Figure~5 shows the
background-subtracted SW count rate between 90 and 180~\AA , while the
lower panel shows the SW hardness ratio, defined as the ratio of the
background-subtracted SW counts from 90--125~\AA \ to that from
125--180~\AA . As in Figure~2, the data are binned into detector ``good
time intervals'' of typically $\approx 1950$~s, and the intervals during
which eclipses occurred are indicated by the circles. The upper panel
demonstrates that the SW light curve is reasonably consistent with the
DS light curve within the errors, while the lower panel reveals a mild
spectral evolution during the observation. A constant $r=a$ fit to the
hardness ratio produces $\rm \chi ^2/dof = 72.8/49=1.49$ with $a=0.77\pm
0.03$, a linear fit $r=a+bt$ produces $\rm \chi ^2/dof = 51.5/48=1.07$
with $b=-0.12\pm 0.03$, and a quadratic fit $r=a+bt+ct^2$ produces $\rm
\chi ^2/dof = 42.7/47=0.91$ with $c=-0.10\pm 0.03$. As measured by the $F$
test, each additional degree of freedom decreases $\chi^2$ sufficiently
to require the additional parameter with $>99\%$ confidence. This
establishes a real but weak correlation of the hardness ratio with the
SW count rate.

To investigate this spectral evolution further, we constructed a
pseudotrailed SW spectrum of OY~Car by concatenating individual spectra
accumulated during the various SW ``good time intervals.'' The result is
shown in Figure~6, where the upper panel shows the spectra as observed,
and the lower panel shows the spectra with the count rate evolution
removed (with the individual spectra normalized to an equal number of
counts between 90 and 180~\AA ). With the count rate evolution removed,
the spectrum is seen to be quite stable throughout the observation: there
are strong emission lines at $\approx 114$, 123, 131, 136, 138, 144, 150,
and 168~\AA , plus what appears to be a weak continuum which peaks at
$\sim 120$~\AA . The only obvious spectral variations are in the relative
normalization of the lines and continuum---the lines being relatively
stronger toward the middle of the observation, and the continuum being
relatively stronger at beginning and end of the observation. These
variations are sufficiently subtle to allow us to assemble a mean
spectrum of OY~Car from the entire observation.

To accomplish this, we maximized the signal-to-noise ratio by combining
data from the SW spectrometer from 90 to 175~\AA \ and data from the MW
spectrometer from 170 to 190~\AA ; the overlap from 170 to 175~\AA \
lends confidence in the emission feature at $\approx 172$~\AA . The
resulting mean {\it EUVE\/} spectrum of OY~Car is shown in Figure~7,
where the SW spectrum is binned to 0.25~\AA , the MW spectrum is
binned to 0.5~\AA \ (roughly half the FWHM of the spectral resolution of
the two spectrometers), and the traces at the bottom of the figure show
the associated $1\, \sigma $ error vectors. If it is not immediately
obvious, a look through the {\it EUVE\/} Stellar Spectral Atlas
\citep{cra97} will convince the reader that the mean {\it EUVE\/}
spectrum of OY~Car is like that of no other class of objects: it lacks
the strong continuum of hot white dwarfs, it lacks the narrow
high-excitation emission lines of late-type stars and RS CVn binaries,
it doesn't look like the spectra of polars, intermediate polars, or even
other dwarf novae. Instead, the spectrum is dominated by broad ($\rm
FWHM\approx 1$~\AA ) emission lines, and the continuum is weak and can be
plausibly described as a superposition of numerous weak emission lines.

In an attempt to quantitatively determine the parameters (wavelength,
width, and flux) of the various emission features, we fit the mean SW
spectrum of OY~Car from 94.5 to 173.5~\AA \ with a number of Gaussians.
Of order 40 Gaussians were required to ``iron out'' the residuals, but
with 316 data points and of order 120 free parameters, the fits tended to
be unstable (minor variations in the initial parameters resulted in major
variations in the fit parameters). Instead, reasonable and stable fits
were obtained with 44 Gaussians with prescribed widths. In one fit, the
Gaussian widths $\sigma $ were fixed at a common value, while in another
they varied with the central wavelength of the Gaussians as $\sigma _i
=[0.212^2+(\lambda _i\Delta v/c)^2]^{1/2}$. The latter choice reproduces
the FWHM of the SW spectrometer if $\Delta v=0$ and accounts for the
broadening of the lines by Doppler processes. The best fits were obtained
with either $\sigma =0.46$~\AA \ (hence the intrinsic $\rm FWHM =
0.96$~\AA ) or $\Delta v = 975~\rm km~s^{-1}$ (hence the intrinsic
$\rm FWHM= 2300~km~s^{-1}$). The parameters from the latter more physical
fit are listed in Table~1.

By comparing the data in Table~1 to the \citet{kel87}; \citet{mew85};
and \citet{ver96} line lists, one sees that a consistent interpretation
of the EUV spectrum of OY~Car can be had with lines from intermediate
ionization stages of cosmically abundant N, O, Ne, Mg, and Fe. The five
brightest lines are identified as follows:
122.9~\AA : \ion{Ne}{6} $2p$--$3d$,
143.5~\AA : \ion{Ne}{5} $2p^2$--$2p3d$,
168.5~\AA : \ion{Fe}{8} $3p^63d$--$3p^53d^2$,
131.3~\AA : \ion{Fe}{8} $3d$--$4f$, and
150.3~\AA : \ion{O}{6} $2s$--$3p$.
These identifications alone significantly constrain the physical nature
of the emitting plasma. If collisionally ionized, its temperature must
lie in the range $\log T({\rm K})\approx 5.4$--5.6. If photoionized, its
ionization parameter $\xi\equiv L/nr^2$ must lie in the range $\log \xi
\approx 0$--5. Because all of the above strong lines are resonance lines,
scattering in a photoionized plasma is the more likely mechanism
responsible for the observed spectrum, but a detailed investigation,
provided in the next two sections, is required to defend this statement.

\section{Spectrum of a Collisionally Ionized Plasma}

To construct a model spectrum of a collisionally ionized plasma, we used
the suite of IDL programs distributed with version 2 of the CHIANTI
atomic database \citep{lan99}. We assumed the elemental abundances of
\citet{all73}; the ionization equilibrium curves of \citet{maz98}; an
absorbing column density $N_{\rm H}=1$--$10\times 10^{19}~\rm
cm^{-2}$; a distance $d=85$ pc, consistent with the values derived by
\citet{woo89} and \citet{bru96}; and a temperature $T=10^5$--$10^6$~K,
incremented in steps of 0.1 in the log. As shown in Figure~8, a model
with $\log T({\rm K})=5.5$, $N_{\rm H}=3.4\times 10^{19}~\rm cm^{-2}$,
and an emission measure $EM\equiv\int n_e^2 dV=3.2\times 10^{55}~\rm
cm^{-3}$ does a reasonable job of reproducing the strongest lines of
\ion{O}{6}, \ion{Ne}{5}, and \ion{Ne}{6}. The strength of the \ion{Fe}{8}
$\lambda 131.1$ emission line can be reproduced if the ionization
fraction $\chi $ of \ion{Fe}{8} is increased by a factor of approximately
two from its value of $\chi=0.4$ at $\log T({\rm K})=5.5$, but none of
\citet{arn85}, \citet{arn92}, or \citet{maz98} have \ion{Fe}{8} peaking
above $\chi= 0.5$, implying that the Fe abundance must be enhanced by a
similar factor to reproduce the data. The \ion{Fe}{8} $\lambda 168.6$
emission line is not reproduced by the model because the data for that
transition is missing from the CHIANTI database. Similarly, the model is
incapable of reproducing the \ion{O}{5} $\lambda\lambda 139.0$, 135.5,
124.6, or 118.0 emission features because the database is missing lines
of \ion{O}{5} shortward of 166~\AA . A more severe problem is the
spectrum of \ion{O}{6}, which is shown shaded in Figure~8: many of the
predicted lines are too strong relative to the data. Particularly
egregious are the $\lambda 115.8$ $2s$--$4p$ resonance line and the
$\lambda 129.8$ $2p$--$4d$ and $\lambda 173.0$ $2p$--$3d$ nonresonance
lines. The $\lambda 173$ line can be reduced in strength by increasing
the absorbing column density, but this makes the problem with the
$\lambda 130$ line worse. Both the $\lambda 173/\lambda 150$ and $\lambda
130/\lambda 150$ line ratios can be decreased by increasing the density,
but values in excess of $10^{14}~\rm cm ^{-3}$ are required. Finally,
optical depth effects would be expected to {\it decrease\/} the strength
of the resonances lines, hence to {\it increase\/} the $\lambda
173/\lambda 150$ and $\lambda 130/\lambda 150$ line ratios, making the
problem worse. Given these problems, it seems unlikely that the EUV
spectrum of OY Car is produced by a collisionally ionized plasma.

\section{Spectrum of Scattering in a Photoionized Plasma}

To produce a scattering model of the EUV spectrum of OY~Car, we need to
specify the EUV spectral energy distribution $F_\lambda $; the wind
geometry, velocity law $v(r)$, and mass-loss rate $\Mdotw $; the
absorbing column density $N_{\rm H}$; the source distance $d$; and the
relevant atomic data. Specifically, in the Sobolev approximation, the
scattered intensity is $F_\lambda\, (1-e^{-\tau_\lambda})$, where the
optical depth $\tau_\lambda = (\pi e^2 /m_{\rm e}c)\, n A \chi
\lambda_{\rm ij} f_{\rm ij}\, (g_{\rm i}/\Sigma g_{\rm i})\,
(dv/dr)^{-1}$, $n$ is the density, $A$ are the abundances, $\chi $ are
the ionization fractions, $dv/dr$ is the wind velocity gradient, and
$\lambda_{\rm ij}$, $f_{\rm ij}$, and $g_{\rm i}$ are respectively the
wavelengths, oscillator strengths, and statistical weights of the various
transitions. We assume that the abundances are as given by \citet{all73},
the atomic data are as given by \citet{ver96}, and the EUV spectral
energy distribution is dominated by that of the boundary layer, which
radiates like a blackbody with temperature $T_{\rm bl}$ and luminosity
$L_{\rm bl} = 4\pi\Rwd ^2f\sigma T_{\rm bl}^4 =4.3\times 10^{34}\, f\,
(T_{\rm bl}/10^5~{\rm K})^4~\rm erg~s^{-1}$, where $\Rwd = 7.8\times
10^8$~cm is the radius appropriate to a $0.7\> \Msun $ white dwarf and
$f$ is the fractional emitting area of the boundary layer.

The observed spectrum can be used to constrain the possible values of
$T_{\rm bl}$, $N_{\rm H}$, and $f$: if the boundary layer temperature
is too high there is too much flux at the shortest wavelengths; if the
absorbing column density is too low there is too much flux at the longest
wavelengths; if the area of the boundary layer is too small it will not
produce the observed flux. These constraints are shown in the upper panel
of Figure~9, where we plot as a function of $T_{\rm bl}$ and $N_{\rm H}$
the value of $f$ required to reproduce the observed flux density.
Consider the shape of the $f=1$ curve, for which the absorbed blackbody
flux distribution ``drapes'' over the emission lines of the observed
spectrum. At low temperatures and absorbing column densities the
constraint is set by the peak flux density of the 96~\AA \ emission line;
as the temperature increases the constraint switches to the 123~\AA \
emission line; at even higher temperatures the constraint switches to the
168~\AA \ emission line. The dotted curves in Figure~9 show where these
transitions take place. Favorable models simultaneously satisfy the 123
and 168~\AA \ constraints, so lie near the upper dotted curve: $T_{\rm
bl}=80$--140 kK, $N_{\rm H}\approx 1.6$--$4\times 10^{19}~\rm cm^{-2}$,
and $f\approx 0.01$--20, hence $L_{\rm bl}\approx 2$--$350\times
10^{33}~\rm erg~s^{-1}$. Models in the upper hatched region of the figure
are excluded because they exceed the Eddington limit of a $0.7\> \Msun $
white dwarf.

The next step is to specify the parameters of the wind. Rather than
attempt a full three-dimensional model, we settle here for a one-zone
model and take fiducial values of the relevant parameters: specifically,
the velocity gradient $dv/dr=3000~\rm km~s^{-1}/10^{10}~cm = 0.03~s^{-1}$
and the density $n=10^{10}~\rm cm^{-3}$ at a distance $r=10^{10}~\rm cm$;
the wind mass-loss rate is then $\Mdotw =4\pi r^2\mu m_{\rm H}nv =
1.3\times 10^{-10}~\rm \Msun~yr^{-1}$. To match the observed widths of
the emission lines, the opacity is distributed as a Gaussian with $\rm
FWHM=0.96$~\AA \ and the resulting model is convolved with a $\rm
FWHM=0.5$~\AA \ Gaussian to account for the spectral resolution of the SW
spectrograph.

To construct a model spectrum within this framework, it is necessary to
specify the shape and intensity of the underlying continuum through the
parameters $T_{\rm bl}$, $N_{\rm H}$, and $f$, and the optical depths of
the lines through the product $nA\chi(dv/dr)^{-1}$. Assuming that $n$
and $(dv/dr)^{-1}$ are as given above, this leaves the three continuum
parameters and the set of products $\chi^\prime \equiv A\chi $ to
specify a given model. This set of models can be constrained further
by considering only the spectrum of \ion{Fe}{8}, which has strong and
reasonably unblended emission lines at 108, 131, 169, and 185~\AA .
Unfortunately, it proves impossible to reproduce the flux in all four of
these emission lines simultaneously, but it is possible to satisfy the
constraints imposed by any combination of three. The locus of points for
Model Sequence~1 satisfying the 108, 131, 169~\AA \ emission line fluxes
is shown in the upper and lower panels of Figure~9 by the lower of the
two bold solid curves, while that for Model Sequence~2 satisfying the
108, 131, 185~\AA \ emission line fluxes is shown by the upper of the two
curves. In each case the optical depth in the \ion{Fe}{8} lines is varied
by varying $\chi^\prime $(\ion{Fe}{8}) = [0.01,0.1,0.3,0.6,1,2,3,6,10]
[$\chi^\prime $(\ion{Fe}{8}) $> 1$ if the Fe abundance is greater
than solar or if the wind density or velocity gradient are greater than
their fiducial values]. At low optical depths and high temperatures
Sequence~1 converges on $T_{\rm bl}=127$~kK and $N_{\rm H} =1.6\times
10^{19}~\rm cm^{-2}$, while Sequence~2 converges on $T_{\rm bl} =104$~kK
and $N_{\rm H}=3.1\times 10^{19}~\rm cm^{-2}$. The trajectory of both
sequences is truncated at high optical depths and low temperatures where
the models first fail to provide enough flux to explain the observed
strengths of other emission lines (the 123~\AA \ emission line for
Sequence~1, the 169~\AA \ emission line for Sequence~2); this happens at
$T_{\rm bl}\approx 100$~kK and $N_{\rm H}\approx 2.3\times 10^{19}~\rm
cm^{-2}$ for Sequence~1, and at $T_{\rm bl}\approx 90$~kK and $N_{\rm H}
\approx 3.5\times 10^{19}~\rm cm^{-2}$ for Sequence~2.

 From among the above model sequences satisfying the \ion{Fe}{8} line
ratios, we present results for five models. Model 1-7a assumes $T_{\rm
bl} =107$~kK, $N_{\rm H}=2.1\times 10^{19}~\rm cm^{-2}$, $f=0.27$, and
$\chi^\prime$(\ion{Fe}{8}) = 3; all other ionization fractions are set
to one. The resulting model spectrum is shown superposed on the data in
the upper panel of Figure~10. By construction, the model matches the
flux in the \ion{Fe}{8} 108, 131, and 169~\AA \ emission lines, but it
overpredicts or underpredicts the flux in other lines. Many of these
deficiencies can be remedied by adjusting the ionization fractions of the
various ions, although blending renders the choices somewhat arbitrary.
Careful examination of the models reveals that the strongest emission
lines are due to Ne, Fe, and O, but lines from Mg, N, Al, Na, and Ar (in
decreasing order of importance) also contribute; the only lines from Si
are from \ion{Si}{5}, and the strongest of its two lines lies at 96.4~\AA
\ in a gap between two observed lines. Given these results, we set the
concentrations of ions of N, Na, Mg, Al, and Ar to 1 and Si to 0. To
minimize the flux of lines at and below 172~\AA , we set the
concentrations of \ion{O}{4}, \ion{Ne}{4}, and \ion{Fe}{9}--\ion{Fe}{11}
to 0. The concentrations of the other ions were adjusted to match the
observed emission line strengths (specifically, the strength of the
lines marked with asterisks in Fig.~10 were adjusted; the strength of the
remaining lines follow according to the strength of the local continuum
and their optical depths). Table~2 lists the resulting model parameters
of this Model 1-7b and the middle panel of Figure~10 shows the resulting
model spectrum. The overall match to the data is reasonably good,
although the flux in the \ion{O}{5} $\lambda 172$ and \ion{Fe}{8}
$\lambda 185$ emission lines are overpredicted and there are a few
lines (e.g., at 128 and 155~\AA ) which these models are simply incapable
of producing. Similar results are obtained for Model 1-5b, for which
$\chi^\prime$(\ion{Fe}{8}) = 1, so it is not shown in Figure~10. The
problem with the \ion{Fe}{8} $\lambda 185$ emission line can be fixed by
switching to the Model Sequence~2. The parameters of Model 2-5b [with
$\chi^\prime$(\ion{Fe}{8}) = 1] and Model 2-3b [with
$\chi^\prime$(\ion{Fe}{8}) = 0.3] are listed in Table~2 and the
Model 2-5b spectrum is shown superposed on the data in the lower panel of
Figure~10. These models match the flux in the \ion{Fe}{8} $\lambda 185$
emission line, but underpredict the flux in the \ion{Fe}{8} $\lambda
169$ line. Both model sequences overpredict the flux in the \ion{O}{5}
$\lambda 172$ emission feature, whose strength is determined by the pair
of \ion{O}{5} emission features at 136 and 139~\AA .

These results demonstrate that it is possible to reproduce the essential
features of the EUV spectrum of OY~Car with a simple model wherein
radiation from the boundary layer is scattered into the line of sight by
the system's accretion disk wind. Given the success of this model, we
supply in Table~1 likely identifications (ions, transitions, wavelengths,
and oscillator strengths) of the observed lines from the \citet{ver96}
list of atomic data for permitted resonance lines. Table~1 and Figure~10
show that the important ions in these models are
\ion{N}{5},
\ion{O}{5}--\ion{O}{6},
\ion{Ne}{5}--\ion{Ne}{7},
\ion{Mg}{4}--\ion{Mg}{6},
\ion{Fe}{6}--\ion{Fe}{8}, and \ion{Fe}{23}.
With the exception of the last ion, the ionization potentials of these
species lie between 100 and 200 eV. The 1950 eV ionization potential of
\ion{Fe}{23} renders the identification of the 133~\AA \ emission feature
suspect, but such a high-excitation [$\log T({\rm K})\approx 7.3$ in
collisional ionization equilibrium] component of the wind cannot be
excluded by the spectrum, because the \ion{Fe}{18}--\ion{Fe}{22} lines
in the SW bandpass are predicted to be weak.

\section{Evaluation of the Radiation Force}

Although it has been known for over two decades that high-$\Mdot $ CVs
have strong winds \citep{hea78, kra81, kla82}, it is still uncertain what
drives the wind. Radiation pressure in spectral lines is implicated by
the similarity of the effective temperatures and P~Cygni profiles of
high-$\Mdot $ CVs with those of early-type stars, but centripetal
acceleration may play a role if the disk is threaded by a large-scale
magnetic field \citep{bla82, can88}. A minimum requirement of any
successful model is that it reproduce the observed wind mass-loss rates
$\Mdotw$ for reasonable mass-accretion rates $\Mdota$ (or, equivalently,
luminosities). Both quantities are uncertain at some level, but
reasonable values are $\Mdota\lax 10^{-8}~\rm \Msun~yr^{-1}$ ($L\lax 20\>
\Lsun $ for a $0.7\> \Msun $ white dwarf) and $\Mdotw\gax 10^{-11}~\rm
\Msun~yr^{-1}$ (where the upper limit is due to the uncertain correction
for the ionization fractions of such species as \ion{C}{4}; a correction
which could easily increase $\Mdotw$ by a factor of 10). Significant
progress has been made recently in models of radiation-driven disk winds
\citep{per97, pro98, fel99a}, but none appear capable of satisfying
these criteria. \citet{per97} obtained $\Mdotw = 2\times 10^{-14}~\rm
\Msun~yr^{-1}$ for a $1\> \Lsun $ accretion disk; \citet{pro98} obtained
$\Mdotw\le 6\times 10^{-12}~\rm \Msun~yr^{-1}$ for a $15\> \Lsun $
accretion disk; \citet{fel99b} obtained $\Mdotw \sim 10^{-12}~\rm
\Msun~yr^{-1}$ for a $10\> \Lsun $ accretion disk. \citet{dre00} argued
that the mass-accretion rates of high-$\Mdot $ CVs must be a higher
than is presently understood for radiation pressure alone to drive the
observed wind mass-loss rates.

Thanks to the edge-on geometry of OY~Car, we can make a reasonably
accurate estimate of the radiation force on the wind, which is simply
the total luminosity of the lines divided by the speed of light: $\Pdot
=L/c$. The uncertainties are the conversion from flux to luminosity and
the contribution from lines out of our bandpass, particularly lines
between 200~\AA \ and the Lyman limit, which are hidden from us by
interstellar absorption. Adding up the flux in the lines listed in
Table~1, the observed {\it EUVE\/} spectrum provides a net radiation
force of $\Pdot\approx 4\times 10^{20}$ dynes, but accounting for
absorption by $\log N_{\rm H}~({\rm cm^{-2}}) =19.2$--19.6 increases
this to $\Pdot\approx 2$--$20\times 10^{21}$ dynes. To determine the
contribution from lines out of our bandpass we need a model of the
intrinsic spectral energy distribution of the boundary layer and
accretion disk and of the ionization state of the wind. To construct a
fiducial model, we assume that the boundary layer radiates like a
blackbody with $T_{\rm bl}=104$ kK and $f=1$ (hence $L_{\rm bl}=5.0
\times 10^{34}~\rm erg~s^{-1}=13\> \Lsun$) and that the disk has the
standard Shakura-Sunyaev temperature profile and radiates locally like
a blackbody with $L_{\rm disk}=L_{\rm bl}$ (hence $\Mdota=1.3\times
10^{-8}~\rm \Msun~yr^{-1}$). If such a system were viewed face-on, it
would have a flux density at 1500~\AA \ of $f_{1500}=3.3\times 10^{-11}\,
(d/85~{\rm pc})^{-2}~\rm erg~cm^{-2}~s^{-1}~\AA ^{-1}$ and a reasonable
absolute visual magnitude of $M_V=5.0$ \citep{war87}. The model spectrum
extends from 1 to 7000~\AA \ (the entire range covered by the
\citealt{ver96} line list), the wavelength bins are set at $3000~\rm
km~s^{-1}$, and all other wind parameters are as given in the previous
section (in particular, $n=10^{10}~\rm cm^{-3}$ and $\Mdotw=1.3\times
10^{-10}~{\rm \Msun~yr^{-1}}=0.01\> \Mdota$). We computed the wind's
ionization state exposed to the radiation fields of the boundary layer
and accretion disk using the atomic physics packages in the current
version of the \citet{ray77} spectral code. In particular, these codes
use the \citet{rei79} photoionization cross sections and low-temperature
dielectronic recombination rates from \citet{nus83}. For simplicity, the
wind temperature is held fixed at 30~kK; varying this quantity simply
shifts the ``horizontal'' positions of the peaks of the various
ionization fractions.

The results of our calculation are shown in Figure~11. The first panel
shows the radiation force as a function of ionization parameter for the
boundary layer alone ({\it dotted curve\/}) and the boundary layer plus
accretion disk ({\it solid curve\/}). More detail is provided for the
latter model in subsequent panels. The second panel shows the
contribution to the radiation force by the various elements: low-Z
elements dominate the radiation force at low ionization parameters; O,
Ne, and Ar dominate at intermediate ionization parameters; and Ne, Mg,
and Fe dominate at high ionization parameters. The contribution by the
various ions of O and Ne as well as the individual contributions of
\ion{O}{6} $\lambda 150$, \ion{Ne}{5} $\lambda 143$, and \ion{Ne}{6}
$\lambda 123$ are shown in the third and fourth panels. That the model is
properly normalized is indicated by the observed values of $\Pdot $ for
these lines based on the fluxes in Table~1 and the assumption that $\log
N_{\rm H}~({\rm cm^{-2}})=19.4\pm 0.2$. The ionization parameter of the
wind is restricted to the range $\log \xi\approx 0$--5 by the weakness
of \ion{O}{4} and \ion{Ne}{4} at the low end of the range and \ion{Ne}{7}
and \ion{Fe}{9} at the high end of the range. To drive the wind to a
terminal velocity of $V_\infty = 3000~\rm km~s^{-1}$, we require $\Pdot
=\Mdotw V_\infty = 2.5\times 10^{24}$ dynes, which is an order of
magnitude larger than the model is capable of producing.

To investigate if other combinations of parameters produce more favorable
results, we tested six other models along the trajectories of the
model sequences from the previous section. With the assumption that
$L_{\rm bl}=4\pi\Rwd^2f\sigma T_{\rm bl}^4=L_{\rm disk}=
G\Mwd\Mdota/2\Rwd $ and that the disk has the standard Shakura-Sunyaev
temperature profile, the maximum disk temperature is $T_{\rm max}=0.49\,
(3f)^{1/4}\, T_{\rm bl}$; requiring that $T_{\rm max}\le T_{\rm bl}$
then limits the range of possible models to those with $f\le 5.8$.
Combined with the previous constraints, such models have EUV spectra
which are dominated by the boundary layer and have net luminosities in
the range $L=2.6\times 10^{34}$--$1.3\times 10^{36}~\rm
erg~s^{-1}=6.8$--$340\> \Lsun $. Wind densities (mass-loss rates) were
taken from the results of the detailed model fits of the previous
section. The resulting parameters for the six models (plus the above
fiducial model) are listed in Table~3 and the results of the wind
momentum calculations are shown in Figure~12. Only the models at the
extreme ends of the two model sequences have $\Pdot/\Mdotw V_\infty$
above unity, hence are capable of driving the required mass-loss rates.
As shown in Table~3, however, these models are unrealistically bright
in the UV ($f_{1500}\approx 1\times 10^{-10}~\rm erg~cm^{-2}~s^{-1}~\AA
^{-1}$) and optical ($M_V\approx 4$), and so are not likely realistic
models of OY~Car. Based on these results, we conclude that radiation
pressure falls roughly an order of magnitude short of driving the wind of
OY~Car in superoutburst.

\section{Discussion and Summary}

Because of the longer time on source, the larger number of eclipses
observed, and the higher count rate in the detector, the {\it EUVE\/}
observations discussed here significantly strengthen previous reports that
there is little or no eclipse by the secondary of the EUV emission region
of OY~Car in superoutburst. The mean {\it EUVE\/} spectrum extends from
70 to 190~\AA \ and contains broad ($\rm FWHM\approx 1$~\AA ) emission
lines of \ion{N}{5}, \ion{O}{5}--\ion{O}{6}, \ion{Ne}{5}--\ion{Ne}{7},
\ion{Mg}{4}--\ion{Mg}{6}, \ion{Fe}{6}--\ion{Fe}{8}, and possibly
\ion{Fe}{23}. The gross details of the spectrum are reproduced by the
spectrum of a collisionally ionized plasma with a temperature $T\approx
320$~kK, an absorbing column density $N_{\rm H}\approx 3.4\times
10^{19}~\rm cm^{-2}$, and an emission measure $EM\approx 3.2\times
10^{55}~\rm cm^{-3}$, but such a model fails to reproduce the observed
spectrum in detail. Better fits are obtained with a model (similar
to that of Seyfert 2 galaxies) wherein radiation from the boundary
layer and accretion disk is scattered into the line of sight by the
system's cool ($T\sim 30$ kK) photoionized accretion disk wind. The
likely range of acceptable parameters for the boundary layer are $T_{\rm
bl}\approx 90$--130 kK and $N_{\rm H}\approx 1.6$--$3.5\times 10^{19}~\rm
cm^{-2}$. This range for the absorbing column density is typical of the
interstellar column densities of nearby CVs beyond the Local Bubble
\citep{mau88, pol90, lon96}, while the range of boundary layer
temperatures is significantly less than that of SS~Cyg \citep[$T_{\rm
bl} \approx 230$--350~kK;][]{mau95} but comparable to that of U~Gem
\citep[$T_{\rm bl}= 110$--140~kK;][]{lon96} and VW~Hyi \citep[$T_{\rm
bl}\approx 120$~kK;][]{mau96}. The corresponding range of the boundary
layer luminosity is less clear because it is possible to trade off the
intensity of the EUV continuum for the wind optical depth (i.e., the
wind mass-loss rate). A lower limit to the fractional emitting area of
the boundary layer is $f\approx 0.1$, and a lower limit to the boundary
layer luminosity is $L_{\rm bl}\approx 1\times 10^{34}~\rm erg~s^{-1}
\approx 3\> \Lsun $. Adding a ``minimal'' Shakura-Sunyaev accretion
disk\footnote{The observed values of $L_{\rm disk}/L_{\rm bl}$ range
from $\approx 2$ for U~Gem \citep{lon96}, $\gax 10$ for SS~Cyg
\citep{mau95}, and $\sim 20$ for VW~Hyi \citep{mau96}.} with $L_{\rm
disk}=G\Mwd \Mdota/2\Rwd =L_{\rm bl}=4\pi\Rwd^2f\sigma T_{\rm bl}^4$
provides further constrains. Requiring that the maximum disk temperature
be less than or equal to the boundary layer temperature requires that
$f\le 5.8$ and $L_{\rm bl}\le 6.5\times 10^{35}~\rm erg~s^{-1}\approx
170\> \Lsun $. Adding reasonable brightness constraints in the UV
[$f_{1500}\lax 1\times 10^{-10}\, (d/85~{\rm pc})^{-2}~\rm
erg~cm^{-2}~s^{-1}~\AA^{-1}$] and optical ($M_V\gax 5$) is equivalent
to the requirement that the mass-accretion rate $\Mdota\lax 10^{-8}~\rm
\Msun~yr^{-1}$, consistent with observations of other high-$\Mdot $ CVs,
hence $L_{\rm bl}\lax 4\times 10^{34}~\rm erg~s^{-1}\approx 10\> \Lsun $.
Such models have wind mass-loss rates of $\Mdotw\lax 10^{-10}~{\rm
\Msun~yr^{-1}}\approx 0.01\, \Mdota $ and fall an order of magnitude
short of driving the wind by radiation pressure alone. An alternative is
that the wind of OY~Car is driven by a combination of radiation pressure
and magnetic forces (a magnetocentrifugal wind). Such winds have been
investigated recently by \citet{pro00}.

\acknowledgments

The observations described here could not have been accomplished without
the alert provided by M.\ Mattiazzo and the optical data provided by M.\
Mattiazzo, N.\ Suntzeff, M.\ Keane, P.\ Nelson, R.\ Stubbings, T.\ Cragg,
and J.\ Herr. Rapid response to the outburst was facilitated by J.\
Mattei, Director of the AAVSO; F.\ Bateson, Director of the VSS/RASNZ;
and VSNET. Members of the AAVSO and RASNZ provided the visual magnitude
estimates from which the optical light curve was assembled. Special
thanks go to W.\ Walker for the optical photometry shown in Figure~3.
The {\it EUVE\/} target-of-opportunity observations were made possible by
the rapid response of {\it EUVE\/} Acting Deputy Director J.\ Vallerga,
{\it EUVE\/} Science Planner B.\ Roberts, the staff of the {\it EUVE\/}
Science Operations Center at CEA, and the Flight Operations Team at
Goddard Space Flight Center. K.~Dere kindly provided assistance with
CHIANTI. The manuscript was improved by the comments and suggestions of
the referee, K.~Long. C.~W.~M.'s contribution to this work was performed
under the auspices of the U.S.\ Department of Energy by University of
California Lawrence Livermore National Laboratory under contract No.
W-7405-Eng-48.

\clearpage % force page break

% Table 1
%---------------------------------------------------------

\begin{deluxetable}{ccccccc}
\small
\tablecolumns{7}
\tablewidth{0pc}
\tablenum{1}
\tablecaption{Line Parameters and Identifications\label{tab1}}
\tablehead{%
\multicolumn{2}{c}{Line Parameters} &
\colhead{} &
\multicolumn{4}{c}{Line Identifications} \\
\cline{1-2}
\cline{4-7} \\
\colhead{Wavelength} &
\colhead{Flux} &
\colhead{} &
\colhead{Wavelength} &
\colhead{} &
\colhead{} &
\colhead{Oscillator} \\
\colhead{(\AA )} &
\colhead{($10^{-13}~\rm erg~cm^{-2}~s^{-1}$)} &
\colhead{} &
\colhead{(\AA )} &
\colhead{Ion} &
\colhead{Transition} &
\colhead{Strength}}
\startdata
$ 95.59\pm 0.08$&  $1.0\pm 0.2$& & \phn95.446&  \ion{Mg}{6}&
$2p^3$--$2p^23d$&      1.26E$+$00\\
$ 97.64\pm 0.08$&  $1.2\pm 0.2$& & \phn97.495&  \ion{Ne}{7}&
$2s^2$--$2s3p$&        4.86E$-$01\\*
                &              & & \phn98.215&  \ion{Ne}{6}&  $2p$--$4d$&
1.22E$-$01\\*
                &              & & \phn98.477&  \ion{Fe}{8}&  $3d$--$6f$&
1.05E$-$01\\
$100.09\pm 0.13$&  $0.7\pm 0.2$& & \phn99.688&  \ion{O }{6}&  $2s$--$6p$&
1.69E$-$02\\
$101.61\pm 0.11$&  $0.8\pm 0.2$& & 101.309&  \ion{Ne}{6}&
$2s^22p$--$2s2p3p$&       2.78E$-$02\\*
                &              & & 101.501&  \ion{Ne}{6}&
$2s^22p$--$2s2p3p$&       1.97E$-$02\\
$105.08\pm 0.09$&  $1.0\pm 0.2$& & 103.999&  \ion{Mg}{5}&
$2p^4$--$2p^34s$&         1.56E$-$01\\*
                &              & & 104.035&  \ion{Mg}{5}&
$2p^4$--$2p^34d$&         1.56E$-$01\\*
                &              & & 104.813&  \ion{O }{6}&  $2s$--$5p$&
3.20E$-$02\\
$108.05\pm 0.07$&  $1.4\pm 0.2$& & 107.993&  \ion{Fe}{8}&  $3d$--$5f$&
1.97E$-$01\\
$109.66\pm 0.07$&  $1.4\pm 0.2$& & 109.411&  \ion{Ne}{6}&
$2s^22p$--$2s2p3p$&       2.92E$-$02\\
$111.41\pm 0.05$&  $2.6\pm 0.2$& & 110.924&  \ion{Mg}{5}&
$2p^4$--$2p^33d$&         1.91E$-$01\\*
                &              & & 111.147&  \ion{Ne}{6}&
$2s^22p$--$2s2p3p$&       1.68E$-$01\\*
                &              & & 111.668&  \ion{Mg}{6}&
$2p^3$--$2p^23s$&         1.24E$-$01\\*
                &              & & 111.675&  \ion{Fe}{7}&  $3d^2$--$3d7f$&
8.09E$-$02\\
$112.81\pm 0.08$&  $1.3\pm 0.2$& & 112.479&  \ion{Fe}{8}&
$3p^63d$--$3p^53d4s$&     9.30E$-$02\\*
                &              & & 113.293&  \ion{Fe}{8}&
$3p^63d$--$3p^53d4s$&     5.45E$-$02\\
$114.53\pm 0.03$&  $4.2\pm 0.2$& & 113.812&  \ion{Mg}{5}&
$2p^4$--$2p^33d$&         9.69E$-$02\\*
                &              & & 114.135&  \ion{Mg}{5}&
$2p^4$--$2p^33d$&         2.87E$-$01\\*
                &              & & 114.200&  \ion{Ne}{6}&
$2s^22p$--$2s2p3p$&       1.04E$-$01\\*
                &              & & 114.358&  \ion{O }{5}&  $2s^2$--$2s8p$&
1.02E$-$02\\*
                &              & & 114.887&  \ion{Mg}{5}&
$2p^4$--$2p^33d$&         2.43E$-$01\\
$116.20\pm 0.06$&  $2.2\pm 0.2$& & 115.824&  \ion{O }{6}&  $2s$--$4p$&
7.41E$-$02\\*
                &              & & 116.161&  \ion{O }{5}&  $2s^2$--$2s7p$&
1.39E$-$02\\*
                &              & & 116.773&  \ion{Fe}{8}&
$3p^63d$--$3p^53d4s$&     1.11E$-$01\\
$117.60\pm 0.06$&  $2.4\pm 0.2$& & 117.269&  \ion{Fe}{7}&  $3d^2$--$3d6f$&
5.31E$-$02\\*
                &              & & 118.000&  \ion{O }{5}&  $2s^2$--$2p4d$&
5.68E$-$03\\
$119.28\pm 0.04$&  $3.9\pm 0.3$& & 118.911&  \ion{Ne}{5}&  $2p^2$--$2p4d$&
1.78E$-$01\\*
                &              & & 119.047&  \ion{Fe}{8}&
$3p^63d$--$3p^53d4s$&     4.25E$-$02\\*
                &              & & 119.102&  \ion{O }{5}&  $2s^2$--$2s6p$&
2.95E$-$02\\
$120.56\pm 0.24$&  $0.7\pm 0.3$& & 120.250&  \ion{Fe}{7}&  $3d^2$--$3d6f$&
1.34E$-$01\\
$121.59\pm 0.15$&  $1.7\pm 0.3$& & 121.781&  \ion{Mg}{5}&
$2p^4$--$2p^33d$&         2.59E$-$01\\
$122.87\pm 0.03$&  $9.4\pm 0.4$& & 122.619&  \ion{Ne}{6}&  $2p$--$3d$&
5.71E$-$01\\*
                &              & & 123.340&  \ion{Mg}{4}&
$2p^5$--$2p^45d$&         9.07E$-$02\\
$124.56\pm 0.11$&  $1.7\pm 0.3$& & 124.616&  \ion{O }{5}&  $2s^2$--$2s5p$&
5.00E$-$02\\
$125.67\pm 0.06$&  $3.4\pm 0.3$& & 125.245&  \ion{Na}{5}&
$2p^3$--$2p^23d$&         1.11E$+$00\\
$127.49\pm 0.08$&  $3.0\pm 0.4$& & 127.179&  \ion{Fe}{7}&
$3p^63d^2$--$3p^53d^24s$& 1.18E$-$01\\
$128.40\pm 0.15$&  $1.8\pm 0.4$& & 128.581&  \ion{Fe}{7}&  $3d^2$--$3d5f$&
1.21E$-$01\\
$129.98\pm 0.10$&  $2.2\pm 0.3$& & 129.729&  \ion{Al}{4}&
$2p^6$--$2p^53d$&         7.78E$-$01\\*
                &              & & 129.925&  \ion{Mg}{4}&
$2p^5$--$2p^44d$&         5.09E$-$02\\*
                &              & & 130.027&  \ion{Mg}{4}&
$2p^5$--$2p^44d$&         1.65E$-$01\\*
                &              & & 130.374&  \ion{Fe}{7}&  $3d^2$--$3d5f$&
2.63E$-$01\\
$131.33\pm 0.04$&  $7.8\pm 0.4$& & 131.121&  \ion{Fe}{8}&  $3d$--$4f$&
5.68E$-$01\\
$132.65\pm 0.06$&  $4.6\pm 0.4$& & 132.007&  \ion{Ne}{5}&
$2s^22p^2$--$2s2p^23p$&   1.05E$-$01\\*
                &              & & 132.830&  \ion{Fe}{23}& $2s^2$--$2s2p$&
1.52E$-$01\\*
                &              & & 132.941&  \ion{Mg}{4}&
$2p^5$--$2p^43d$&         4.50E$-$02\\
$134.17\pm 0.15$&  $1.2\pm 0.3$& & \nodata & \nodata & \nodata & \nodata \\
$136.07\pm 0.05$&  $4.0\pm 0.3$& & 135.523&  \ion{O }{5}&  $2s^2$--$2s4p$&
7.77E$-$02\\
$137.68\pm 0.14$&  $2.3\pm 0.5$& & \nodata & \nodata & \nodata & \nodata \\
$138.65\pm 0.08$&  $4.2\pm 0.5$& & 138.555&  \ion{Ne}{6}&  $2p$--$3s$&
2.83E$-$02\\*
                &              & & 139.029&  \ion{O }{5}&  $2s^2$--$2p3d$&
5.79E$-$02\\
$140.77\pm 0.13$&  $1.8\pm 0.3$& & 140.296&  \ion{Mg}{4}&
$2p^5$--$2p^43d$&         1.42E$-$01\\*
                &              & & 140.357&  \ion{N }{5}&  $2s$--$6p$&
1.59E$-$02\\*
                &              & & 140.571&  \ion{Mg}{4}&
$2p^5$--$2p^43d$&         7.90E$-$02\\*
                &              & & 140.652&  \ion{Mg}{4}&
$2p^5$--$2p^43d$&         1.65E$-$01\\
$142.44\pm 0.19$&  $1.9\pm 0.5$& & 142.615&  \ion{Ne}{5}&  $2p^2$--$2p3d$&
2.27E$-$01\\
$143.51\pm 0.05$&  $8.8\pm 0.6$& & 143.316&  \ion{Ne}{5}&  $2p^2$--$2p3d$&
6.73E$-$01\\
$145.51\pm 0.18$&  $1.5\pm 0.3$& &  \nodata & \nodata & \nodata & \nodata \\
$147.14\pm 0.11$&  $2.8\pm 0.4$& & 146.790&  \ion{Mg}{4}&
$2p^5$--$2p^43d$&         8.47E$-$02\\*
                &              & & 147.153&  \ion{Mg}{4}&
$2p^5$--$2p^43d$&         3.28E$-$01\\*
                &              & & 147.425&  \ion{N }{5}&  $2s$--$5p$&
3.00E$-$02\\
$148.85\pm 0.17$&  $2.0\pm 0.4$& & 148.843&  \ion{Ne}{4}&
$2p^3$--$2p^24d$&         2.89E$-$01\\
$150.33\pm 0.08$&  $6.1\pm 0.6$& & 150.101&  \ion{O }{6}&  $2s$--$3p$&
2.65E$-$01\\*
                &              & & 150.922&  \ion{Fe}{7}&  $3d^2$--$3d4f$&
6.58E$-$01\\
$151.69\pm 0.11$&  $4.0\pm 0.5$& & 151.682&  \ion{Fe}{7}&  $3d^2$--$3d4f$&
4.61E$-$01\\
$153.37\pm 0.95$&  $0.4\pm 0.4$& & \nodata & \nodata & \nodata & \nodata \\
$154.96\pm 0.19$&  $2.7\pm 0.6$& & \nodata & \nodata & \nodata & \nodata \\
$156.06\pm 0.36$&  $1.2\pm 0.6$& & \nodata & \nodata & \nodata & \nodata \\
$158.44\pm 0.28$&  $1.1\pm 0.4$& & 157.717&  \ion{Ne}{4}&
$2p^3$--$2p^24s$&         4.55E$-$02\\
$160.59\pm 0.33$&  $1.1\pm 0.5$& & \nodata & \nodata & \nodata & \nodata \\
$162.65\pm 0.35$&  $1.1\pm 0.5$& & 162.559&  \ion{N }{5}&  $2s$--$4p$&
6.86E$-$02\\
$165.61\pm 0.12$&  $3.7\pm 0.6$& & 165.530&  \ion{Fe}{6}&
$3p^63d^3$--$3p^53d^4$&   4.97E$-$01\\
$168.53\pm 0.06$&  $8.6\pm 0.7$& & 168.638&  \ion{Fe}{8}&
$3p^63d$--$3p^53d^2$&     5.70E$-$01\\
$171.54\pm 0.14$&  $4.0\pm 0.7$& & 171.073&  \ion{Fe}{9}&
$3p^6$--$3p^53d$&         3.05E$+$00\\*
                &              & & 171.873&  \ion{Mg}{4}&
$2p^5$--$2p^43s$&         7.12E$-$02\\*
                &              & & 172.169&  \ion{O }{5}&  $2s^2$--$2s3p$&
3.95E$-$01\\*
                &              & & 172.567&  \ion{Ne}{4}&
$2p^3$--$2p^23d$&         9.27E$-$01\\*
\enddata
\end{deluxetable}

%---------------------------------------------------------

% Table 2
%---------------------------------------------------------

\begin{deluxetable}{lcccc}
\small
\tablecolumns{5}
\tablewidth{0pc}
\tablenum{2}
\tablecaption{Scattering Model Parameters\label{tab2}}
\tablehead{%
\colhead{Parameter} &
\colhead{Model 1-7b} &
\colhead{Model 1-5b} &
\colhead{Model 2-5b} &
\colhead{Model 2-3b}}
\startdata
\hbox to 1.1in{$T_{\rm bl}$ (kK)\leaders\hbox to 5pt{\hss.\hss}\hfil}&
107&   120&  98.6&  102\\
\hbox to 1.1in{$N_{\rm H}~\rm (cm^{-2})$\leaders\hbox to
5pt{\hss.\hss}\hfil}&       $2.1\times 10^{19}$& $1.8\times 10^{19}$&
$3.2\times 10^{19}$& $3.1\times 10^{19}$\\
\hbox to 1.1in{$f$\leaders\hbox to 5pt{\hss.\hss}\hfil}&
0.27&  0.17& 3.4& 6.5\\
\hbox to 1.1in{$L_{\rm bl}~\rm (erg~s^{-1})$\leaders\hbox to
5pt{\hss.\hss}\hfil}&   $1.5\times 10^{34}$& $1.5\times 10^{34}$&
$1.4\times 10^{35}$& $3.1\times 10^{35}$\\
\hbox to 1.1in{$\chi^\prime $(N)           \leaders\hbox to
5pt{\hss.\hss}\hfil}& 1.0\phn&  0.3\phn& 0.3\phn& 0.1\phn\\
\hbox to 1.1in{$\chi^\prime $(\ion{O}{5 }) \leaders\hbox to
5pt{\hss.\hss}\hfil}& 0.5\phn&  0.2\phn& 0.19   & 0.06   \\
\hbox to 1.1in{$\chi^\prime $(\ion{O}{6 }) \leaders\hbox to
5pt{\hss.\hss}\hfil}& 0.2\phn&  0.08   & 0.09   & 0.03   \\
\hbox to 1.1in{$\chi^\prime $(\ion{Ne}{5 })\leaders\hbox to
5pt{\hss.\hss}\hfil}& 1.0\phn&  0.32   & 0.37   & 0.11   \\
\hbox to 1.1in{$\chi^\prime $(\ion{Ne}{6 })\leaders\hbox to
5pt{\hss.\hss}\hfil}& 2.5\phn&  0.72   & 0.68   & 0.19   \\
\hbox to 1.1in{$\chi^\prime $(\ion{Ne}{7 })\leaders\hbox to
5pt{\hss.\hss}\hfil}& 0.5\phn&  0.1\phn& 0.1\phn& 0.02   \\
\hbox to 1.1in{$\chi^\prime $(Na)          \leaders\hbox to
5pt{\hss.\hss}\hfil}& 1.0\phn&  0.3\phn& 0.3\phn& 0.1\phn\\
\hbox to 1.1in{$\chi^\prime $(Mg)          \leaders\hbox to
5pt{\hss.\hss}\hfil}& 1.0\phn&  0.3\phn& 0.3\phn& 0.1\phn\\
\hbox to 1.1in{$\chi^\prime $(Al)          \leaders\hbox to
5pt{\hss.\hss}\hfil}& 1.0\phn&  0.3\phn& 0.3\phn& 0.1\phn\\
\hbox to 1.1in{$\chi^\prime $(Ar)          \leaders\hbox to
5pt{\hss.\hss}\hfil}& 1.0\phn&  0.3\phn& 0.3\phn& 0.1\phn\\
\hbox to 1.1in{$\chi^\prime $(\ion{Fe}{6 })\leaders\hbox to
5pt{\hss.\hss}\hfil}& 1.0\phn&  0.3\phn& 0.6\phn& 0.2\phn\\
\hbox to 1.1in{$\chi^\prime $(\ion{Fe}{7 })\leaders\hbox to
5pt{\hss.\hss}\hfil}& 0.6\phn&  0.4\phn& 0.3\phn& 0.1\phn\\
\hbox to 1.1in{$\chi^\prime $(\ion{Fe}{8 })\leaders\hbox to
5pt{\hss.\hss}\hfil}& 3.0\phn&  1.0\phn& 1.0\phn& 0.3\phn\\
\hbox to 1.1in{$\chi^\prime $(\ion{Fe}{23})\leaders\hbox to
5pt{\hss.\hss}\hfil}& 4.0\phn&  1.5\phn& 1.5\phn& 0.5\phn\\
\enddata
\end{deluxetable}

%---------------------------------------------------------

% Table 3 (prints sideways)
%---------------------------------------------------------
%\input table3.tex
%---------------------------------------------------------

\clearpage % force page break

\addtocounter{page}{1}		% leave space for Table 3

% References
%---------------------------------------------------------

\clearpage % force page break

% Figure captions
%---------------------------------------------------------

\begin{figure}
\begin{center}
\includegraphics[width=2.5in]{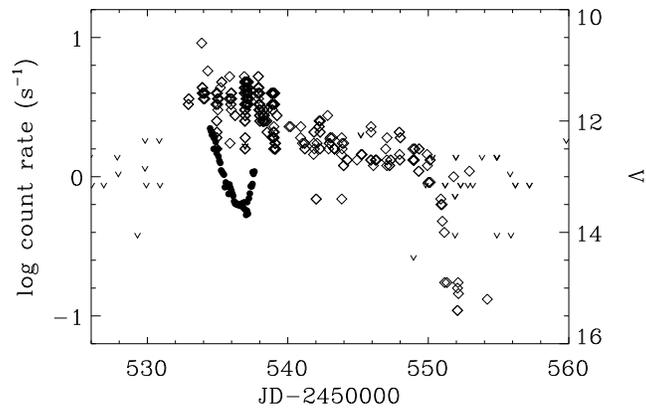}
\end{center}
\vspace*{1 in}
\figcaption[figure1.ps]
{DS count rate and AAVSO and RASNZ optical light curves. DS measurements
are shown by the filled circles; optical measurements and upper limits
are shown by the diamonds and carets, respectively. Optical eclipses are
excluded from this presentation by restricting the binary phases to $0.1
\le\phi\le 0.9$.
\label{fig1}}
\end{figure}
\clearpage

\begin{figure}
\begin{center}
\includegraphics[width=2.75in]{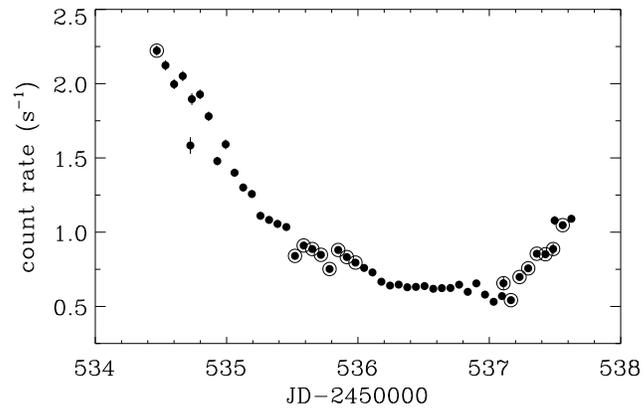}
\end{center}
\vspace*{1 in}
\figcaption[figure2.ps]
{DS count rate light curve. Orbits containing eclipses of the white
dwarf are marked with circles. Error bars are typically smaller than the
symbols.
\label{fig2}}
\end{figure}
\clearpage

\begin{figure}
\begin{center}
\includegraphics[width=5.00in]{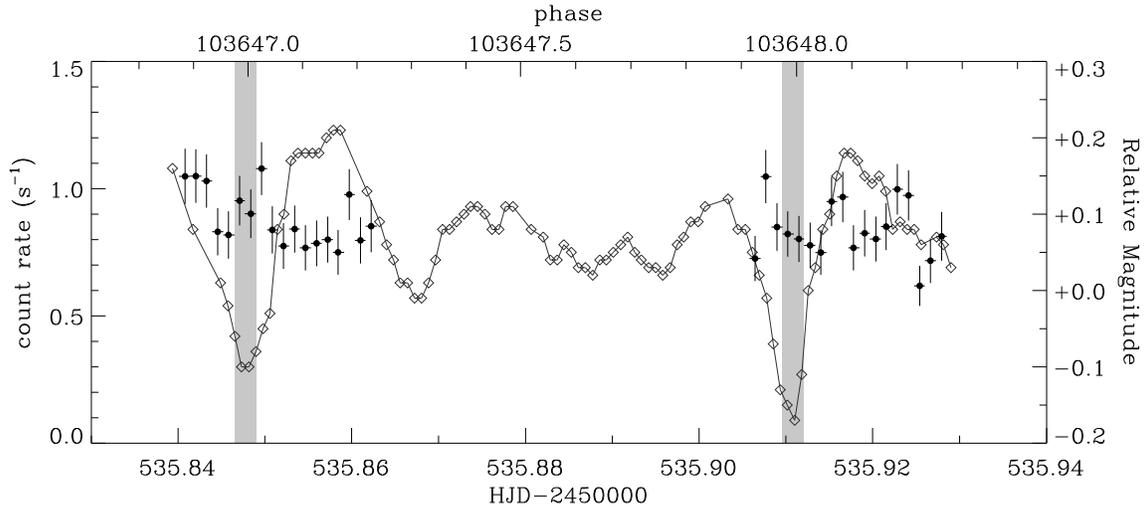}
\end{center}
\vspace*{1 in}
\figcaption[figure3.ps]
{DS count rate and optical light curves for the interval $\rm HJD-2450000
=535.84$--535.93. DS measurements are shown by the filled circles with
error bars, relative optical measurements are shown by the diamonds
connected by straight lines. Binary phase is relative to the ephemeris
of \citet{pra99a}. The grey stripes denote the phase intervals when the
white dwarf is eclipsed by the secondary ($-0.02\le \phi \le +0.02$).
\label{fig3}}
\end{figure}
\clearpage

\begin{figure}
\begin{center}
\includegraphics[width=2.75in]{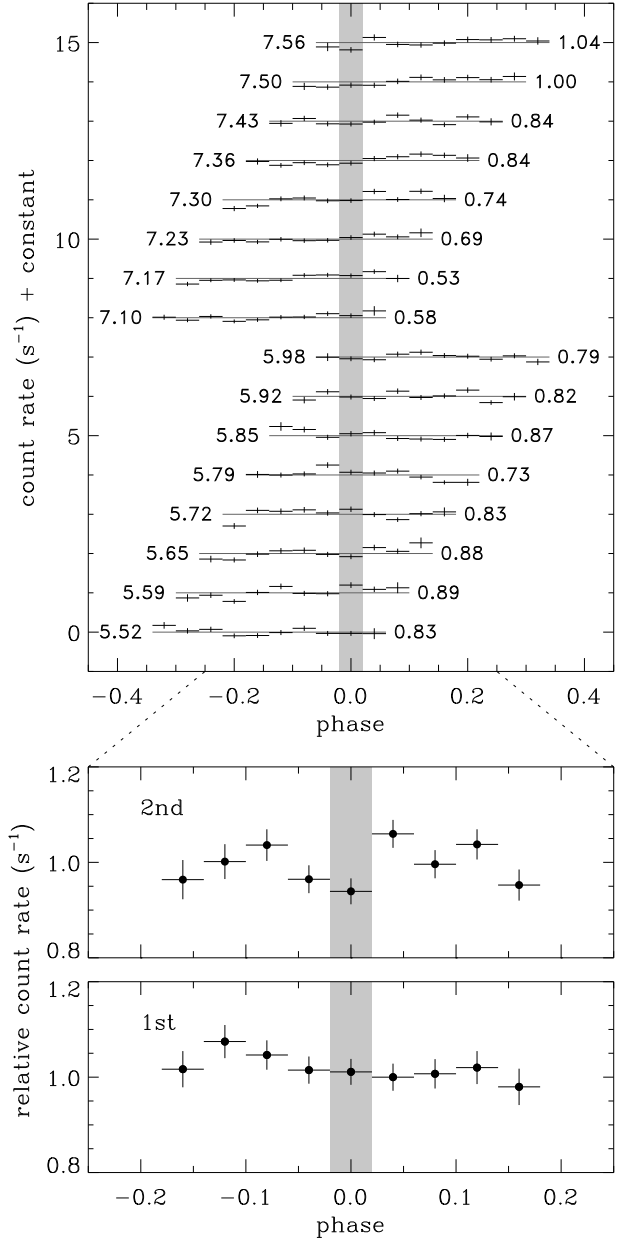}
\end{center}
\vspace*{0.5 in}
\figcaption[figure4.ps]
{{\it Upper panel:\/} DS count rate light curves from individual {\it
EUVE\/} orbits. Label to the right of each curve is the date ($\rm
HJD-2450530$) of the midpoint of the orbit, while that to the right is
the weighted mean count rate $\bar{c}$. Each successive orbit of data
is offset vertically by $n-\bar{c}$, where $n=0$, 1, \ldots , 15.
{\it Lower panels:\/} Mean relative DS light curves from the first and
second groups of eclipse intervals shown in the upper panel. The grey
stripes denote the phase interval when the white dwarf is eclipsed by
the secondary ($-0.02\le \phi \le +0.02$).
\label{fig4}}
\end{figure}
\clearpage

\begin{figure}
\begin{center}
\includegraphics[width=2.75in]{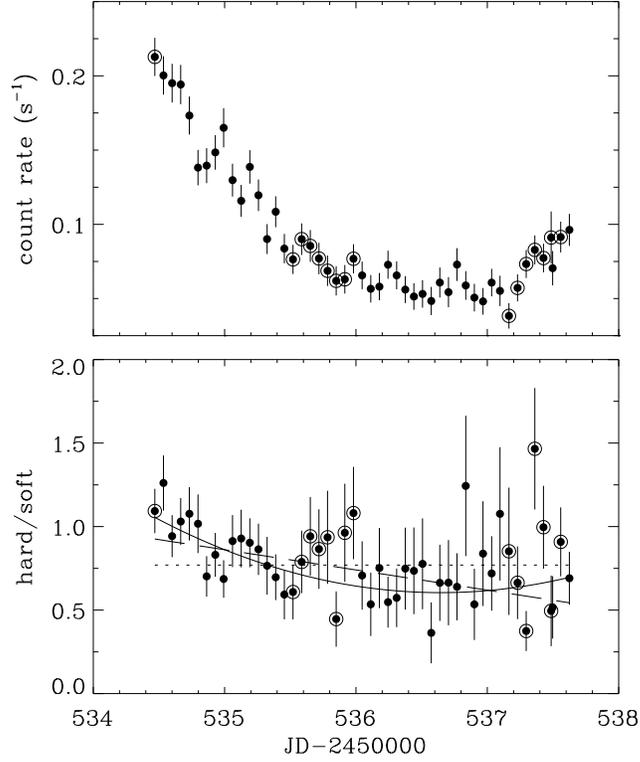}
\end{center}
\vspace*{1 in}
\figcaption[figure5.ps]
{SW count rate ({\it upper panel\/}) and hardness ratio ({\it lower
panel\/}) light curves. Hardness ratio is defined as the ratio of
the background-subtracted SW counts from 90--125~\AA \ to that from
125--180~\AA . Constant, linear, and quadratic fits of the hardness
ratio are indicated respectively by the dotted, dashed, and solid
curves.
\label{fig5}}
\end{figure}
\clearpage

\begin{figure}
\begin{center}
\includegraphics[width=5.50in]{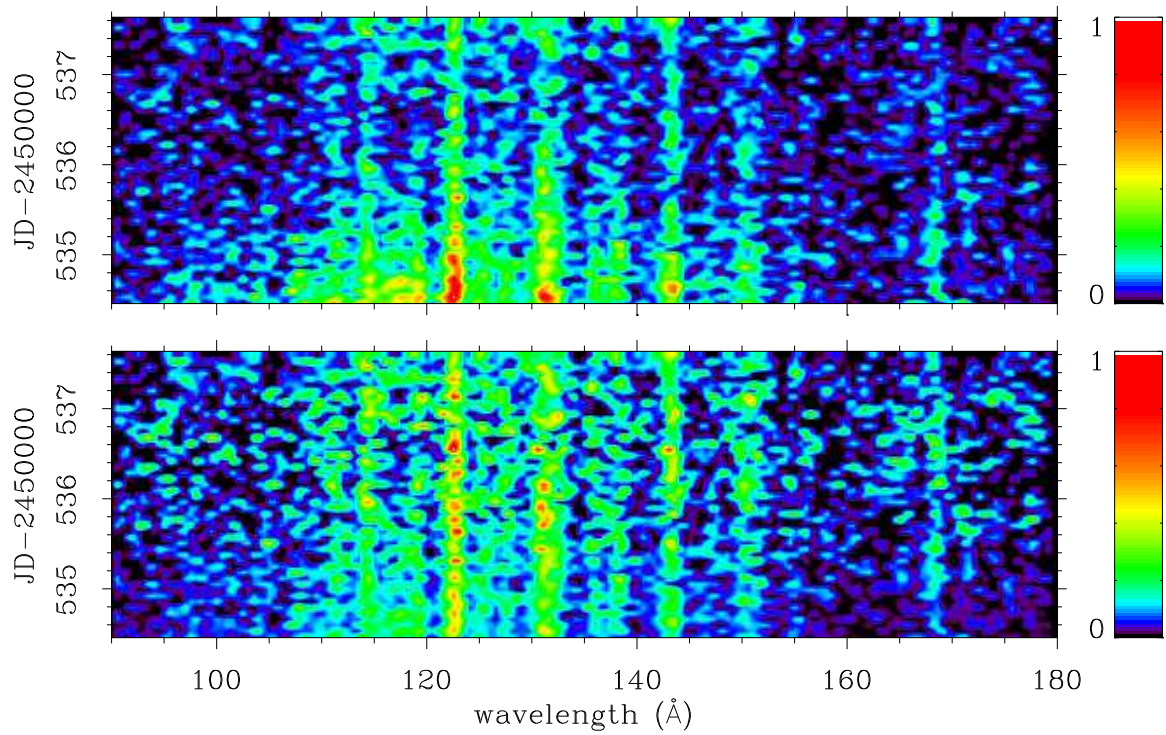}
\end{center}
\vspace*{1 in}
\figcaption[figure6.ps]
{Pseudotrailed SW spectrum. {\it Upper panel:\/} spectra as observed.
{\it Lower panel:\/} spectra normalized to an equal number of counts
between 90 and 180~\AA .
\label{fig6}}
\end{figure}
\clearpage

\begin{figure}
\begin{center}
\includegraphics[width=5.50in]{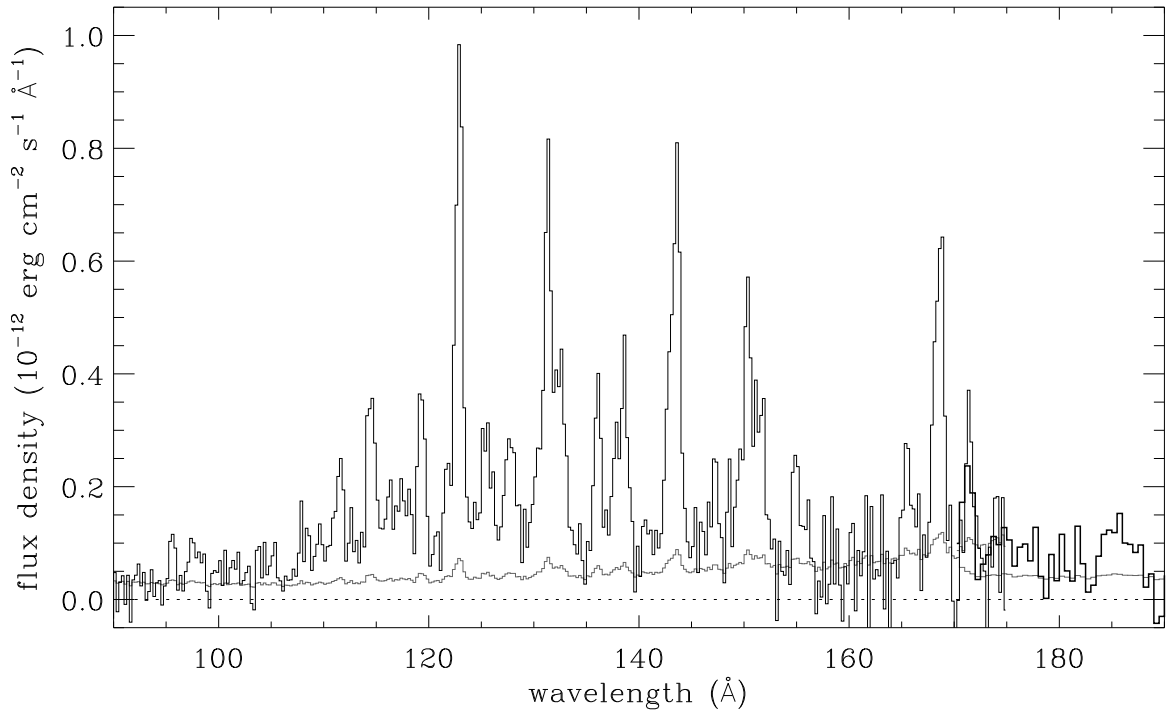}
\end{center}
\vspace*{1 in}
\figcaption[figure7.ps]
{Mean SW and MW spectra. The SW spectrum ends at 175~\AA\ while the MW
spectrum begins at 170~\AA . Traces at the bottom of the figure are
the associated $1\, \sigma $ error vectors.
\label{fig7}}
\end{figure}
\clearpage

\begin{figure}
\begin{center}
\includegraphics[width=5.50in]{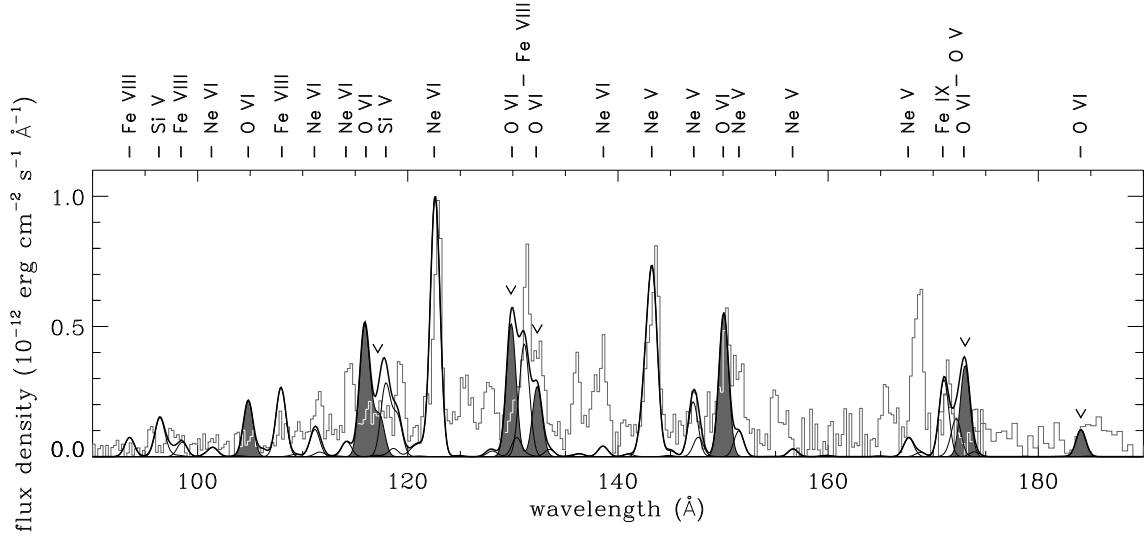}
\end{center}
\vspace*{1 in}
\figcaption[figure8.ps]
{CHIANTI collisional plasma model of the mean {\it EUVE\/} spectrum of
OY~Car with $\log T({\rm K})=5.5$, $N_{\rm H}=3.4\times 10^{19}~\rm
cm^{-2}$, $EM =3.2\times 10^{55}~\rm cm^{-3}$, and convolved with a
$\rm FWHM=1.08$~\AA \ Gaussian. The observed spectrum is shown by the
grey histogram, the net model spectrum by the thick solid curve, and
the spectra of the various ions by the thin solid curves. The \ion{O}{6}
model spectrum is shaded and its nonresonance lines are indicated by
carets.
\label{fig8}}
\end{figure}
\clearpage

\begin{figure}
\begin{center}
\includegraphics[width=2.75in]{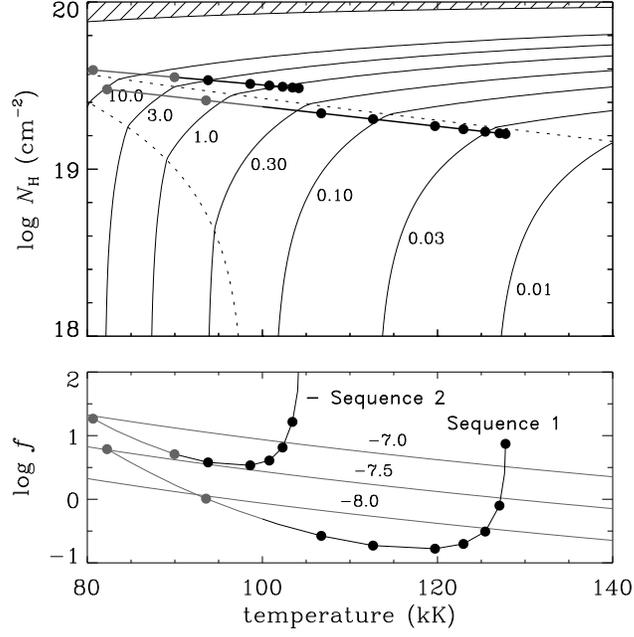}
\end{center}
\vspace*{1 in}
\figcaption[figure9.ps]
{{\it Upper panel:\/} Fractional emitting area $f$ for $\tau=\infty$
as a function of temperature $T_{\rm bl}$ and absorbing column density
$N_{\rm H}$. Favorable continuum models lie near the upper dotted curve.
Models in the hatched region exceed the Eddington limit. Solid lines
are the locus of points satisfying the \ion{Fe}{8} emission line
flux constraints. {\it Lower panel:\/} Fractional emitting area $f$
satisfying the \ion{Fe}{8} emission line flux constraints with
$\chi^\prime $(\ion{Fe}{8}) = [0.01,0.1,0.3,0.6,1,2,3,6,10]. Grey
curves are contours of constant luminosity $L_{\rm bl}= 4\pi\Rwd
^2f\sigma T_{\rm bl}^4 = G\Mwd\Mdota/2\Rwd$ for $\Mdota= 10^{-8}$,
$10^{-7.5}$, and $10^{-7}~\rm \Msun~yr^{-1}$.
\label{fig9}}
\end{figure}
\clearpage

\begin{figure}
\begin{center}
\includegraphics[width=5.50in]{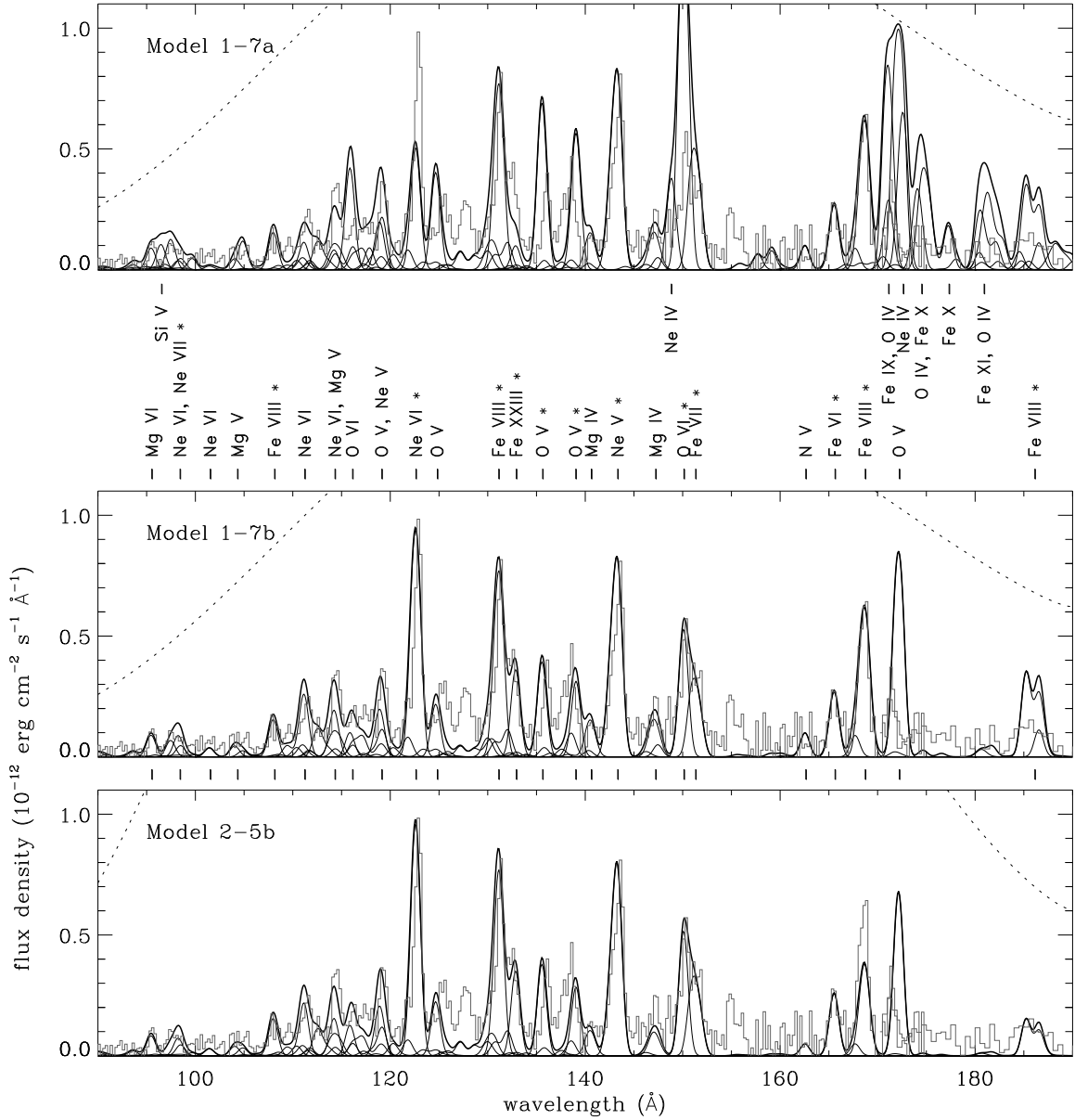}
\end{center}
\vspace*{0.5 in}
\figcaption[figure10.ps]
{Representative scattering models of the mean {\it EUVE\/} spectrum of
OY~Car. The observed spectrum is shown by the grey histogram, the
absorbed blackbody continua by the dotted curves, the net model spectra
by the thick solid curves, and the spectra of the various ions by the
thin solid curves. The lines marked with asterisks were adjusted to
achieve the fits shown. Refer to Table~2 for model parameters.
\label{fig10}}
\end{figure}
\clearpage

\begin{figure}
\begin{center}
\includegraphics[width=2.75in]{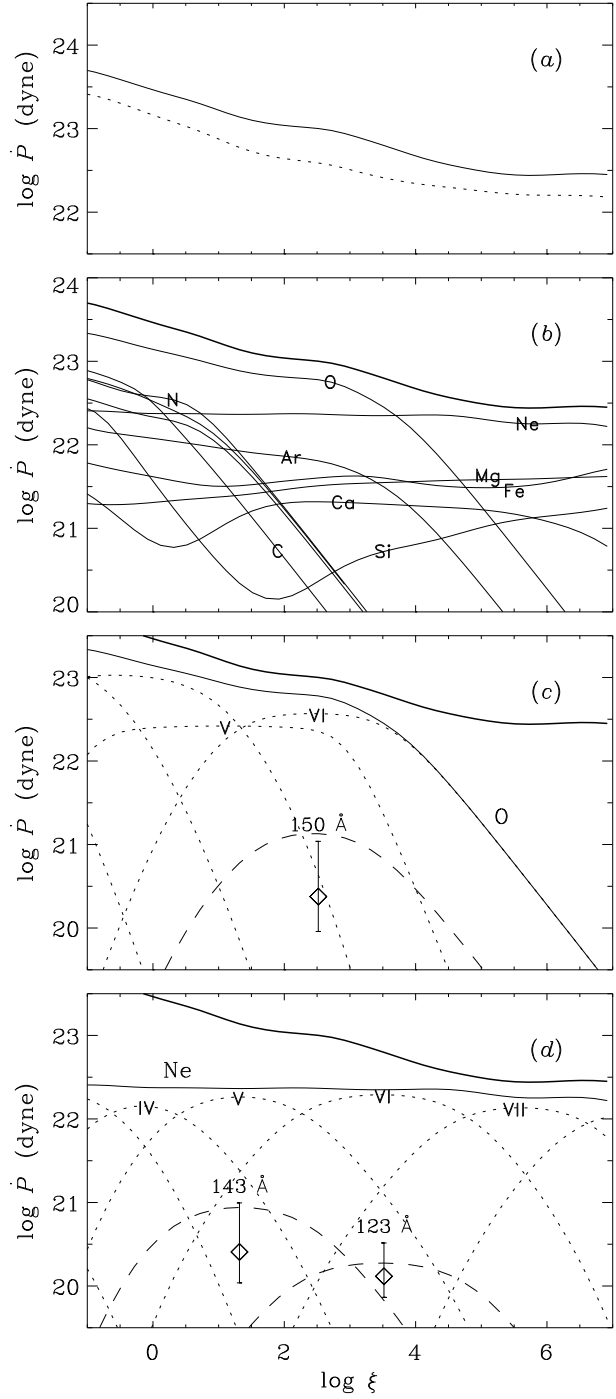}
\end{center}
\vspace*{0.25 in}
\figcaption[figure11.ps]
{Radiation force $\Pdot = L/c$ as a function of ionization parameter $\xi
$. ({\it a\/}) Total force from the boundary layer alone with $T_{\rm
bl}=104$ kK and $f=1$ (hence $L_{\rm bl}=13\> \Lsun $) ({\it dotted
curve\/}) and the boundary layer plus accretion disk with $L_{\rm
disk}=L_{\rm bl}$ ({\it solid curve\/}). ({\it b\/}) Contributions from
the various elements, ({\it c\/}) ions of O, and ({\it d\/}) ions of Ne
for the model with $L_{\rm disk}=L_{\rm bl}$. Curves labeled ``150~\AA
'', ``143~\AA '', and ``123~\AA '' are respectively the contributions of
\ion{O}{6} $\lambda 150$, \ion{Ne}{5} $\lambda 143$, and \ion{Ne}{6}
$\lambda 123$. Data points are the observed values of these lines
assuming $d=85$ pc and $\log N_{\rm H}~({\rm cm^{-2}}) =19.4\pm 0.2$.
\label{fig11}}
\end{figure}
\clearpage

\begin{figure}
\begin{center}
\includegraphics[width=2.75in]{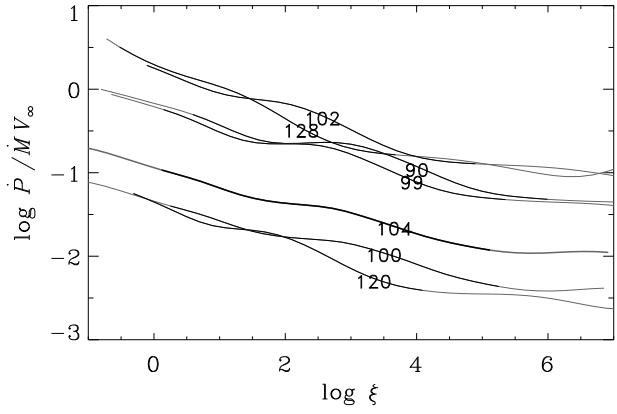}
\end{center}
\vspace*{1 in}
\figcaption[figure12.ps]
{Radiation force $\Pdot$ relative to the wind momentum $\Mdotw V_\infty$
as a function of ionization parameter $\xi $. Various curves are labeled
with the boundary layer temperature $T_{\rm bl}$ (kK) and are dark where
the models likely apply: where the contributions of \ion{O}{6} $\lambda
150$, \ion{Ne}{5} $\lambda 143$, or \ion{Ne}{6} $\lambda 123$ are within
a half a magnitude of their peak values.
\label{fig12}}
\end{figure}

\end{document}